# Confinement Highlights the Different Electrical Transport Mechanisms Prevailing in Conducting Polymers

*Sukanya Das,[1] Anil Kumar,[2] and K. S. Narayan[1,\*]*

[1]Chemistry and Physics of Materials Unit and School of Advanced Materials, Jawaharlal Nehru Centre for Advanced Scientific Research, Bengaluru- 560064, India.

[2]Department of Chemistry, Indian Institute of Technology Bombay, Powai, Mumbai- 400076, India.

**ABSTRACT**

**We study the differences in electrical charge transport dynamics of the conductivity enhancement of poly(3,4-ethylenedioxythiophene) (PEDOT) derivatives under geometrical confinement. The results of polymer blend poly(3,4-ethylene dioxythiophene):poly(styrenesulfonate) and a polymer-monomer blend, poly(3,4-ethylenedioxythiophene):tosylate, highlight the role of dopants and processing conditions of these systems under confinement. The prevailing transport length scales in confined geometry of characteristic dimensions originate from varying disorder in these polymer systems. These observable differences in two different PEDOTs introduced by molecular level reorganization can be utilized to tune conducting polymer systems for efficient electrical and thermoelectric properties. The electrical conductivity $\sigma$ of the polymer system, which is a function of the electronic structure at molecular level and a connectivity parameter, has been probed in cylindrical-alumina nanoscaffolds of various channel diameters, at different frequencies $\omega$ and temperatures T. The observations also emphasize the role of disorder in these conducting polymer systems.**

## INTRODUCTION

Tuning of the structural organization at different length-scales in conducting polymer (CP) blends has helped achieve significant milestones in the electrical transport properties. The charge carriers in these CPs traverse different electrical transport regimes ranging from molecular scales to microscopic and mesoscopic length scales. Structural-tuning of the appropriate length scale is a key-material design tool. Since CPs are inherently disordered and consists of polymer chains with different characteristic lengths and defect-distribution, micro-structural modifications can alter the transport mechanism significantly.[1-6]

With numerous conformational degrees of freedom, the charge transport dynamics vary in different regimes of $\pi$-stacked molecular planes. Localization of charges occurs due to the absence of polymer chain ordering over long distances and crystal symmetry. This restricts diffusion of charges and limits charge transport mechanism to hopping between localized sites, in the framework of Anderson model.[7] Differences in electronic delocalization length scale within molecular stacks contribute to the structural disorder in the conjugated polymers. The coexistence of delocalized and localized states in high molecular weight polymers can be interpreted in terms of transport routes and characteristic barriers which act as rate-limiting processes[8]. The transport effectively occurs via intermediate order-disorder pathways of the semicrystalline-amorphous microstructural domains.[9] The properties in the ordered domains set the upper limit for the charge transport in such heterogeneous microstructures.[10] However, the enhanced $\sigma$ has also been observed to originate from effective $\pi$-$\pi$ stacking at molecular levels as opposed to contributions solely from improved-crystallinity.[11] The rate-limiting factor for charge transport dynamics described by the structure property relation of CPs is thus system-specific.

Among the CPs, poly(3,4-ethylenedioxythiophene) PEDOT emerges as a versatile candidate with significantly high electrical conductance, thermal and air stability, simple synthesis process, energy storage capacity, optical transparency and biocompatibility.[12-19] PEDOT:PSS is an archetypical CP blend where extensive studies have been carried out. The importance of secondary doping processes has been demonstrated for improving σ such as solvent-additive introduction, ionic liquids, surfactant treatment, acid treatment, to name a few.[12] Enhanced σ in PEDOT:PSS has been explained widely in terms of polymer chain alignment emphasizing conformational change from worm-like chains to linear structures, [20, 21] electronic and ionic phase separation of PEDOT and PSS, respectively[22], PEDOT-PSS complex grain size[23] , removal of PSS from PEDOT:PSS grains[24] and ordering of PEDOT nanocrystals[9, 25]. From only a few S/cm to ∼ 6000 S/cm, with PEDOT single nanocrystals reaching up to 8797 S/cm[26], the electrical σ of PEDOT is close to that of conventional metals. The other derivative of PEDOT where significant electrical and thermoelectric properties have been observed is PEDOT:Tos, in which a controlled oxidation level during the process of EDOT polymerization in presence of iron(III)-tosylate has significantly increased the σ.[5, 27, 28] With the presence of small non-hindering tosylate counterions and absence of long insulating polymer PSS, PEDOT:Tos is more ordered and exhibits metallic transport signatures[16, 29]. The wide range of electronic properties prevailing in systems with a common PEDOT backbone is interesting and poses fundamental questions about factors controlling the properties. Specifically, we probe these two contrasting PEDOT systems: PEDOT:Tos and PEDOT:PSS, respectively, in which the in-situ polymerization in confinement leads to characteristic chain organization and compare it to the microstructure which is developed upon insertion of another all-polymer blend in the nano cavities. PEDOT chains within such narrow cylindrical cavities offer a test-bed for understanding the structural disorder and its role in electrical transport.

Straight-through hollow alumina nanopores provide rigid-inert, geometrically-ordered confinement for these systems. Confined CPs have been previously shown to undergo the general feature of chain alignment accompanied by an enhancement in the conductance. In our previous work[30] we showed that the transverse (nanopore axial) conductivity $\sigma_{transverse}$ of PEDOT:PSS increases as geometrical confinement increases to 20 nm. In this report, we implement these studies for the in-situ grown PEDOT:Tos that provides an entirely distinct dopant environment and highlight the contrasting transport mechanisms upon comparison with that of PEDOT:PSS. The results can be inferred in terms of correlating the polymer chain alignment and role of macromolecular organizational processes for $\sigma$ to the structural disorder in these low-dimensional templates. These representatives of PEDOT highlight the variation introduced by different dopants on the disorder affecting the extent of delocalization. Previous studies of PEDOT:PSS have shown that the fraction of polarons and bipolarons formed depends on the dopant concentration while in the case of PEDOT:Tos it is predominantly bipolaronic in character.[16] We focus on the domain of chain alignment that maps to the order-disorder boundaries of the polymer blend in the constrained space. It is expected that the molecular rearrangement within nanochannels will result in different ordered phases and extent of their interconnects in the disordered matrix. We investigate the changes in heterogeneous media under confinement and study the response due to small signal electrical perturbation over a wide frequency spectrum and temperature range. The insight gained by confining these systems highlights the order-disorder boundaries and charge dynamics of CPs.

The confinement effect studies are implemented using alumina templates in a film form adhered to a conducting (ITO) substrate. The parallel array of cylindrical pores chosen for the studies are of diameter 20 nm, 50 nm, 80 nm and 100 nm in the alumina matrix (200 nm thick).

The deposition of PEDOT:Tos in the nanochannels involved in-situ polymerization of the monomer within the alumina scaffold (details in sec. 1 of the **Supplementary Material [44], [31-33]**).

Structural studies using X-ray scattering of solution cast polymerized PEDOT:Tos films have indicated a paracrystalline state with a sizable anisotropy, where the dopant anions form distinct planes that alternate with stacks of the polymer chains[29]. These films exhibit $\sigma$ with a metallic behavior in-plane and a dielectric behavior out-of plane of the films. In the context of PEDOT:Tos grown in confined anodized aluminium oxide (AAO) nanopores, it is expected that crystallization leads to a preferential orientation. This effect is characteristic of anisotropic crystals, an orientation with the fast-growing crystallographic direction preferring to be along the long-axis of the nanopore. An unimpeded growth parallel to the pore axis, especially below a characteristic radial-confinement length scale can be expected. These factors then enable enhancement in transverse conductance of the polymer systems.

**RESULTS AND DISCUSSIONS**

*(a) Dc transport $\sigma(T)$*

Four-probe measurements are carried out to obtain the dc lateral conductivity $\sigma_{lateral}$ of PEDOT:Tos films of different thicknesses on a variety of substrates. $\sigma_{lateral}$ (**Figure 1a**) is obtained to be in the range $\sim 500$ ($\pm 100$) S/cm. These values and $\sigma_{lateral}(T)$ are similar to the reported semi-metallic trend of PEDOT:Tos[16]. The $\sigma_{transverse}$ measured between bottom-ITO electrode and top-Au electrode of the films are observed to be lower than $\sigma_{lateral}$ by nearly four orders of magnitude $\sim 10^{-2}$ S/cm. It should be pointed out that $\sigma_{transverse}$ inhomogeneity has been observed in the case of vapor phase polymerized films[34]. The characteristic of $\sigma_{transverse} <$ $\sigma_{lateral}$ can be attributed to the anisotropy of the crystalline segment and the favoured inter-connected links between the growing crystallites lateral to the amorphous film surface. This scenario is interestingly altered for the nanopore-confined CPs.

The $\sigma$ of single PEDOT:Tos nanochannel in the AAO template can be obtained from $\sigma_{transverse}$ measurement with the knowledge of the pore-density and this is possible since PEDOT:Tos fills the alumina pores forming a uniform film. A complete pore-filling has been ensured by individually conducting nanopore in Conductive Atomic Force Microscopy (CAFM) studies in the following section and extensive methods mentioned in sec. 2 of the **Supplementary Material [44].** In the present case of the polymer blend confined in alumina nanochannels, $\sigma_{transverse}$ ($\approx$ 24$\pm$18 S/cm) increases by nearly three orders of magnitude as compared to as-cast bulk films ($\approx 10^{-2}$ S/cm) (**Figure 1b**). It should be emphasized that the $\sigma_{dc}$ is obtained after ascertaining the linear I-V characteristics (0 - 0.8 V) and $\sigma_{transverse}$ represents $\sigma$ of a single nanochannel, obtained by normalizing cumulative response from $10^9$ pores/cm$^2$. The electrical response from templated film of $\sim 10^9$ parallel, identical and well-separated polymer nanopillars can be taken to be an aggregate measure of $\sigma$ from each nanopore. The pore diameter dependence indicates that $\sigma_{transverse}$ of the polymers in the nanopores is not a simple scaled magnitude of the bulk-$\sigma_{transverse}$. The pore-induced reconfiguration of the CP chains leads to a significant enhancement of $\sigma$ and this dramatic non-linear scaling gets evident when the geometrical confinement dimensions get comparable to charge transport length-scales.

The transport properties of these conducting polymer systems are generally characterized by their degree of disorder which is ascertained from $\sigma_{dc}(T)$ and the resistivity ratio $\rho_r = \frac{\rho(0\ K)}{\rho(300\ K)}$.[17, 35, 36] Higher value of $\rho_r$ corresponds to higher levels of disorder. $\sigma_{dc}(T)$ of PEDOT:Tos and PEDOT:PSS are studied for different nanochannel dimensions over a range of 10 K< T < 340 K (**Figure 1c** and sec. 5 of the **Supplementary Material [44]**). The $\sigma_{dc}$ values for all PEDOT:Tos samples can be extrapolated to a finite magnitude as T approaches zero. The nonzero conductivity indicates the presence of charges in Fermi level and the possible conduction pathways without any thermal activation. $\sigma_T/\sigma_{300K}$ versus T is plotted for both PEDOT:Tos and PEDOT:PSS films in **Figure 1d**. In comparison to the PEDOT:PSS samples with a high degree of disorder ($\rho_r \sim 10^3$), the PEDOT:Tos films reveal ordered phases with $\rho_r$ ~1. The conductivity, $\sigma_{dc}(T)$ of confined PEDOT:Tos within different channel-dimensions shows a negative coefficient of temperature ($\frac{d\sigma}{dT} < 0$) over a wide temperature range above a transition temperature (T>$T_M$). In the range T < $T_M$, $\frac{d\sigma}{dT} > 0$ and $\sigma_{dc}(T)$ exhibits a weak temperature dependence with a negligible decrease in the magnitude of σ as T is decreased to 10 K. $T_M$ indicated in **Figure 1d(inset)** is observed to be a function of the nanochannel diameters (sec. 5 of the **Supplementary Material [44]**). Additionally, the high-resolution transmission electron microscopy (HRTEM) images of PEDOT:Tos films show the presence of crystalline PEDOT domains surrounded by amorphous regions (sec. 6 of the **Supplementary Material [44]**). These results, however, necessitate a detailed analysis for the structural study of polymer chains in the confined channels.

In comparison, the PEDOT:PSS nanochannels (**Figure 1c**) show positive coefficient of temperature throughout the range 10 K< T< 340 K. PEDOT:PSS have been extensively reported to exhibit variable range hopping (VRH) type transport or nearest neighbour hopping mechanism depending on the film anisotropy.[17, 37] The analysis in terms of reduced activation energy is provided in sec. 5.1 of the **Supplementary Material [44].** The decreasing trend of $\sigma_{dc}(T)$ for PEDOT:Tos in the range 10 K < T < $T_M$ may suggest some of the residual resistances that are typically observed in metallic samples. The nonzero conductivity as T →0, improved order parameter in the existing disordered matrix and $\frac{d\sigma}{dT} < 0$ over wide temperature range suggest a Sheng conduction where fluctuation induced tunneling[2, 5, 38] occurs between conducting domains separated by thin insulating ones.

*(b) Single nanochannel transport*

The partially filled PEDOT:Tos nanochannels are locally probed using conductive atomic force microscopy by applying a constant bias voltage between conducting cantilever tip and ITO. It is possible to measure the current magnitude across single channels using the Pt/Ir coated nanoconducting tip which has radius of curvature less than the used pore diameter. The single nanochannel measurements (sec. 2 and 4 of the **Supplementary Material [44]**) of PEDOT:Tos reveal a trend similar to that obtained from macroscopic measurements. $I_{20\ nm} > I_{50\ nm} > I_{100\ nm} > I_{bulk}$ for a constant voltage V has been observed using the same CAFM tip which clearly confirms the nanoconfinement effect. $\sigma_{20\ nm}$ is three orders of magnitude higher than $\sigma_{bulk.}$ The analysis process for arriving at conductivity values are similar to our previous work[30] and the relatively minor discrepancy in $\sigma$ from the two methods reveals the maximum uniformity of PEDOT:Tos across the $10^9$ parallel-identical channels.

**Figure 2a** shows surface image of a top section of 50 nm PEDOT:Tos nanochannels, a portion of which is removed manually by tape exfoliation (details mentioned in sec. 2 of the **Supplementary Material [44]**)**.** The partial removal of top PEDOT layer exposes the polymer filled nanochannels (marked with dotted line in **Figure 2a**) whose corresponding surface topography and phase images have been shown in **Figure 2b,d**, respectively**.** The non-exfoliated region has a surface roughness and phase shown in **Figure 2c,e.** The guided lines in **Figure 2a** reveal the exposed cross sectional surface scan of 50 nm nanochannels that provides the evidence of the conformal polymer filling within the nanochannels. The details of the complete exfoliation of AAO-PEDOT template from the ITO surface is comprehensively presented in sec. 2 of the **Supplementary Material [44]**. These results indicate a complete filling of polymer along the length axis of nanochannels. **Figure 2g,i** are the surface topography of 20 nm PEDOT:Tos/ITO interface and high resolution 700 nm*700 nm scan on only 20 nm PEDOT:Tos region, while **Figure 2h,j** are the corresponding phase images. **Figure 2k,l** are the topography and current images, respectively of 80 nm channels. The surface topography, phase and electric current profiles for the different nanochannel dimensions are shown in sec. 3 and 4 of the **Supplementary Material [44].**

*(c) Ac conduction*

$\sigma(\omega)$ behavior of alumina-nanopore filled polymer provides useful insight into the carrier dynamics of disorderd polymer systems. Changing the dopant environment around the electronic backbone of PEDOT from a long-chain polymer PSS to distribution of small tosylate monomers induces a different length of delocalization of charge carriers and hence the extent of disorder in two polymer blend systems. The prominent highlights of these results (**Figure 3a,b**) are the following: (i) $\sigma(\omega)$ of PEDOT:Tos is independent of $\omega$ in 10 kHz $< \omega <$ 1 MHz over 80 K $<$ T $<$ 340 K range; (ii) In the high $\omega$ region, 1 MHz $< \omega <$ 10 MHz and 80 K $<$ T $<$ 340 K regime, a weak $\omega$-dependence emerges and is described by a power-law frequency dependence of conductivity, $\sigma \sim \omega^s$ where 0 $<$ s $<$ 0.07($\pm$0.035). The variation of exponent s remains nearly constant for all PEDOT:Tos nanochannels ($\sigma(\omega,T)$ of other AAO channels are shown in sec. 7 of the **Supplementary Material [44]**). On the other hand for PEDOT:PSS (**Figure 3c**), the range of exponent is 0 $<$ s $<$ 0.89 ($\pm$0.38) and shows a wide dispersion of $\sigma(\omega)$ over T range. The results of $\sigma(\omega)$ for PEDOT:Tos highlight the weak dependence of $\sigma$ on $\omega$ and T as compared to that of PEDOT:PSS. Normalized $\sigma(\omega)$ is plotted together in **Figure 3d** for 20 nm channels of PEDOT:Tos and PEDOT:PSS. The temperature dependence of onset frequency, $\omega_o(T)$ where $\sigma(\omega_o) = 1.1\sigma_{dc}$ is plotted in **Figure 3d (inset)** for two different PEDOT systems. The correlation length scales as $\lambda$, where $\lambda \propto 1/\omega_o$ corresponds to the path lengths along the conducting polymer network.[39, 40] Observations reveal a temperature independent shorter $\lambda$ associated with PEDOT:Tos and a temperature dependent, significantly larger $\lambda$ for PEDOT:PSS for a temperature variation over 80 K< T < 340 K. The temperature independent s and $\omega_o$ (and $\lambda$) trend of PEDOT:Tos emphasizes the negligible thermal disorder of PEDOT:Tos over PEDOT:PSS. These results strongly relates to the metallicity and the improved order parameter in PEDOT:Tos system.

*(d) Electrical noise analysis*

Electrical noise characterization is studied to probe the microscopic details of local current inhomogeneities over the bulk resistances[41]. Noise analysis of electrical transport can reveal the extent of disorder in the electronic states. Time-dependent fluctuations in current (noise) provide insight into the energetic landscape traversed by charge carriers[42] and can be analysed in the $\omega$ domain as power spectrum density (PSD) response. Noise-fluctuations inversely scale with the conductance in the systems, with PEDOT:PSS exhibiting higher noise amplitudes than PEDOT:Tos channels. This result highlights the higher degree of disorder in PEDOT:PSS.[43] Both the systems exhibited the characteristic $1/f^{\gamma}$ behavior where Hooge's exponent lies in $1.2 < \gamma < 2$ for PEDOT:PSS and $\gamma \approx 2$ for PEDT:Tos. However, the noise characteristics showed the following traits: (i) The $1/f^{\gamma}$ response from the normalized $S_I(\omega)/I^2$ in the case of PEDOT: PSS depended on the dc-current magnitude (controlled by bias) and indicates a dispersive electrical transport behavior [Fig. 4(a)]; (ii) $S_I(\omega)/I^2$ in 153K < T < 400K range for PEDOT:PSS shows a variation of three orders of magnitude [Fig. 4(a) for 100 mV] as compared to a largely bias-independent, T-independent $S_I(\omega)/I^2$ noise response from PEDOT:Tos nanochannels [Fig. 4(b)]; (iii) Sweeping through different bias conditions of 100 mV, 500 mV, 800 mV at different temperatures there is a decrease in slope $|-dS_I/d\omega|$ with increasing bias (inset of **Figure 4a** at T = 313 K). Whereas, $|-dS_I/d\omega|$ is fairly constant for different bias-voltage conditions in PEDOT:Tos (inset of **Figure 4b** at T = 313 K); (iv) Corner frequency, fc variation with T for different dimensions of PEDOT:PSS nanochannels with 20 nm at 100 mV being most dispersive and PEDOT:Tos nanochannels has nearly no T dependence (**Figure 4c**); (v) The range of $\gamma$ for different bias conditions and T-range ($T_{min}$= 153 K and $T_{max}$= 340 K in **Figure 4d**) shows low value of $\gamma \approx 1.2$ for 20 nm channels at high bias V= 800 mV and low T. (Noise-response from other nanochannel dimensions as function of $V_{ext}$ and T are shown in sec. 8 of the **Supplementary Material [44]**). Extent of dispersion at high $\omega$ and low T can also be captured by corner frequency, $f_c$, beyond which an apparent white-noise feature emerges. $f_c$ increases linearly with T for PEDOT:PSS nanochannels, 20 nm at 100 mV being most dispersive (sec. 8 of the **Supplementary Material [44]**). Extent of dispersion at high $\omega$ and low T can also be captured by corner frequency fc, beyond which an apparent white-noise feature emerges. fc increases linearly with T for PEDOT:PSS nanochannels, 20 nm at 100 mV being most dispersive (Sec. 8 of the **Supplemental Material [44]**). The different factors contributing to the measured noise in these systems are quite difficult to resolve. However, a qualitative analogy can be made with impedance studies. The temperature-dependent dispersive behavior of $S_I(\omega)/I^2$ at high $\omega$ in the case of PEDOT: PSS is indicative of a stronger temperature dependence of $\sigma$ as opposed to a weak $\sigma(T)$ behavior in PEDOT:Tos. These characteristic fluctuations in PEDOT:PSS can be correlated to the orientational processes accompanied by a thermal disorder.

**CONCLUSION**

In conclusion, a comparative study of the two CPs, PEDOT:PSS and PEDOT:Tos, confined within nanopores reveals the importance of structural disorder and its role in transport kinetics. The large increase of $\sigma$ within the nanopore scaffold provides a route to enhance the vertical transport of PEDOT films. The increase in $\sigma$ by several orders of magnitude is observed to occur for systems where the polymer emerges from in situ polymerization of the monomer within the nanochannels as compared to polymers inserted from a solution phase into the channels. The frequency domain $\sigma$ and time-domain noise studies provide further insight with $\sigma(\omega)$ in disordered PEDOT:PSS exhibiting a power-law behavior and wide dispersion both over $\omega$ and $T$ , in contrast to the negligibly varying $\sigma(\omega,T)$ and comparatively less disordered PEDOT:Tos. Based on these studies, the confinement template offers some insight into the role of different dopant environments around an electronic backbone.

**ACKNOWLEDGEMENTS**

We acknowledge K.L. Narasimhan and N.S. Vidhyadhiraja for useful discussions. S.D thank A. Sundaresan, D.P. Panda and M. Shrivastava for the use of low temperature dc measurement facility. KSN acknowledges support from DST-JC Bose fellowship.

**FIGURES**

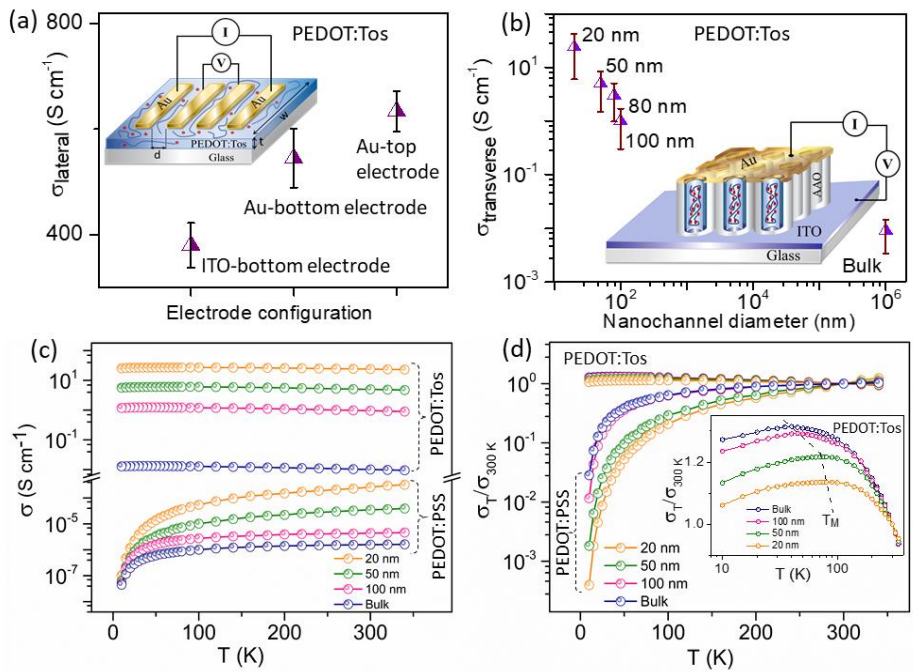

**Figure 1.** Dc conduction of PEDOT:Tos: (a) Lateral dc conductivity, $\sigma_{lateral}$ of PEDOT:Tos films (of thickness 200 nm) for three different configurations of electrode-contact on the film with four-probe method on a glass substrate with patterned-ITO as bottom electrodes, thermally-evaporated bottom-Au electrodes and Au as top electrodes. (b) $\sigma_{transverse}$ of PEDOT:Tos nanochannels between bottom-ITO and top-Au electrodes versus polymer filled alumina nanochannel diameters. Temperature dependent transverse dc conduction: (c) $\sigma(T)$ of PEDOT:Tos and PEDOT:PSS nanochannels in the temperature range 10 K < T < 340 K for different channel dimensions and (d) $\sigma_T/\sigma_{300K}(T)$ (with y-axis in log-scale) for all PEDOT:Tos and PEDOT:PSS channels, inset shows $\sigma_T/\sigma_{300K}(T)$ in linear y-axis for PEDOT:Tos nanochannel diameters with indication line of the transition temperature $T_M$.

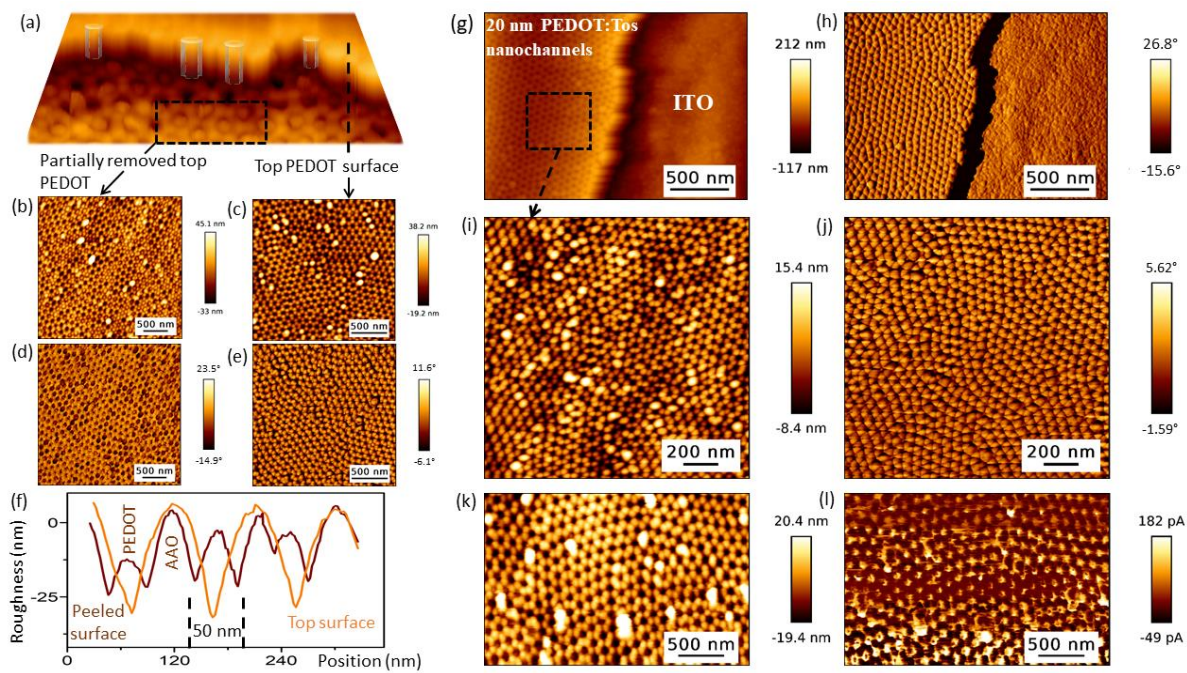

**Figure 2.** Single nanochannel transport. (a) Surface morphology of PEDOT:Tos filled 50 nm AAO surface, a section of which is partially removed as indicated by a dotted square region. The resultant cross-sectional image of the interface shows the exposed PEDOT nanopillars, few of which have been guided with colours. Partial removal of top PEDOT-filled AAO surface shows evidence of PEDOT filling through the length of the channels: (b,d) represents topography and phase, respectively of the dotted square region which comprises of exposed bare alumina and PEDOT filled nanochannels. The alternate bare-AAO and PEDOT channels are also shown in the line scan of surface roughness (f) marked in brown. Line scan of surface roughness(f) marked in orange shows a minima (top concave meniscus of PEDOT adhering to inner surface of nanocylinder formed on annealing) corresponding to each maxima of PEDOT (resultant convex meniscus of PEDOT formed midway a nanocylinder due to peeling of top PEDOT film) in other line scan marked in brown. (c,e) are the respective topography and phase images of top surface of PEDOT filled 50 nm channels. Non-contact mode AFM imaging showing (g,i) topography and (h,j) phase across 20 nm PEDOT:Tos AAO/ITO interface over a scan area of 1.2 μm*1.8 μm and on PEDOT:Tos filled 20 nm AAO surface over a scan area

of 700 nm*700 nm, respectively. Conductive AFM in contact-mode showing (k) topography and (l) current profile of 80 nm PEDOT:Tos nanochannels.

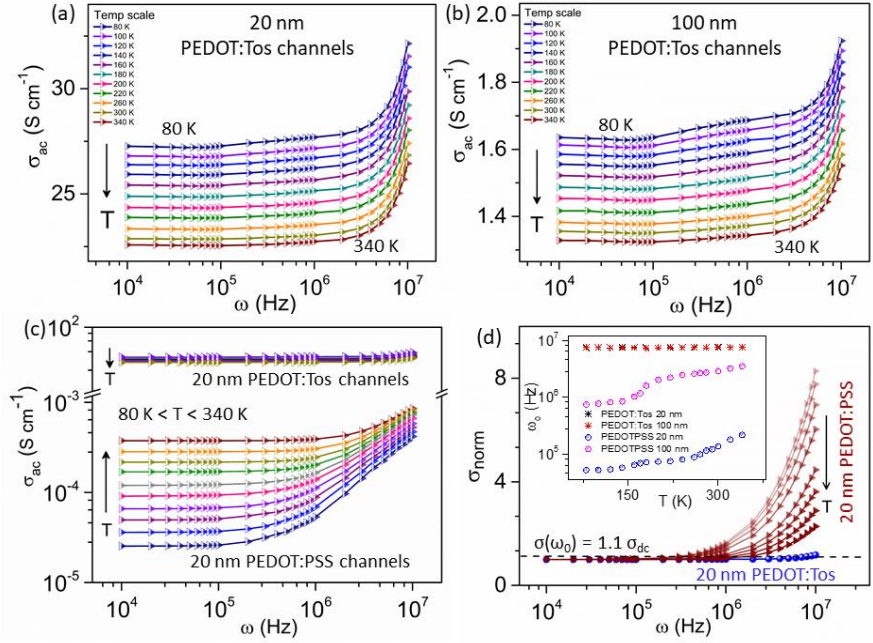

**Figure 3.** Ac conduction in (a) 20 nm and (b) 100 nm PEDOT:Tos channels for temperature range 80 K <T < 340 K. (c) represents the σ(ω,T) behavior for two polymer blends in 20 nm channels: σ$_{PEDOT:PSS}$ shows a wide dispersion over ω and T scales in comparison to precisely unchanged σ$_{PEDOT:Tos}$. (d) Normalized ac conductivity, σ$_{norm}$ of two polymer blends, PEDOT:PSS (increasing color shades of brown in accordance to increasing T scales) and PEDOT:Tos (marked in blue); inset shows the onset frequency ω$_0$(T) for PEDOT:PSS and PEDOT:Tos in the range 80 K < T < 340 K.

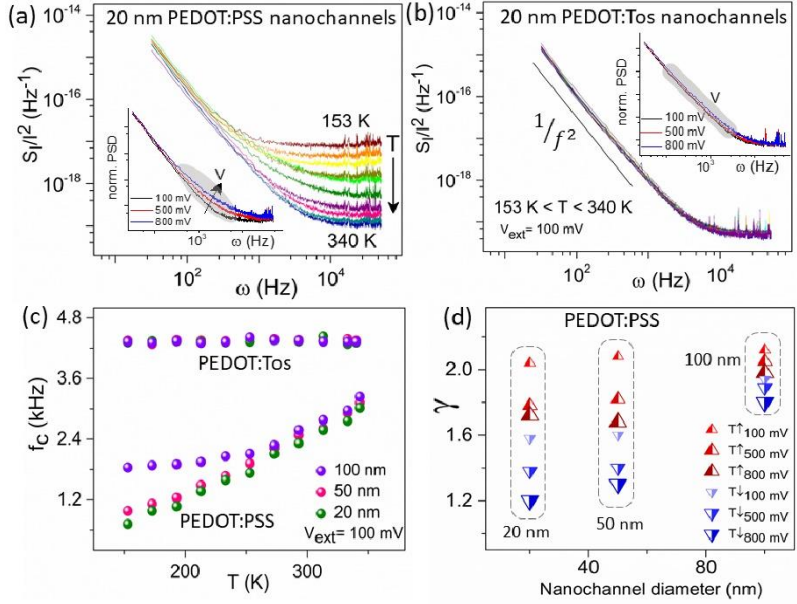

**Figure 4.** Noise studies, $S_I(\omega)/I^2$ in (a) 20 nm PEDOT:PSS nanochannels, (b) 20 nm PEDOT:Tos nanochannels over a temperature range of 153 K < T < 340 K, respectively. Insets of (a) and (b) show the respective normalized PSD with bias conditions. (c) Corner frequency $f_c$ versus T for PEDOT:PSS and PEDOT:Tos channel dimensions. (d) Hooge's exponent, $\gamma$ with different bias conditions at max T ↑ = 340 K (red) and min T ↓ = 153 K (blue) as parameters for 20 nm, 50 nm and 100 nm PEDOT:PSS nanochannels, respectively.

**Supplementary Material**

# Confinement Highlights the Different Electrical Transport Mechanisms Prevailing in Conducting Polymers

**List of contents:**

## 1. Experimental methods

### (a) Materials

Film deposition of PEDOT:Tos involved utilization of monomer EDOT and oxidant solution iron(III)-tosylate. An aqueous dispersion of PEDOT:PSS (PH1000 grade) containing 30 wt.% solid content was procured from Ossila with PEDOT to PSS ratio being 1:2.5. Anodized aluminium oxide (AAO) templates supported on PMMA layer were obtained from commercial sources.

### (b) Sample Preparation

*PEDOT:Tos film preparation:* The alumina membranes (AAO) were transferred onto cleaned ITO glass substrates as outlined in earlier work[1]. These membranes had been used previously in our laboratory to demonstrate polymer-based transistor behavior from individual vertical nanochannels[2]. The standard protocol for chemical polymerization of PEDOT:Tosylate was adopted[3]. The mixture containing solution of EDOT, iron(III)-tosylate in butanol and pyridine was left for stirring for 20 min and then filtered using 0.45 µm PVDF syringe filters. The filtered solution was spin-coated at 1000 rpm on the plasma-treated alumina-ITO substrates. The samples were heated at 363 K for 6 hrs in nitrogen environment and cooled to room temperature. The formation of a yellow-brown coloration of the film was indicative of the presence of $Fe^{2+}(Tos^-)_2$ salt. The film on being repeatedly washed with ultrapure ethanol solution transformed to a blue-colored PEDOT:Tosylate film with no traces of iron-tosylate salts (Energy Dispersive X-Ray Analysis, EDX results in **Figure S1**). A complete filling of the pores was accomplished through the layer-by-layer deposition of PEDOT:Tos, wherein the hydrophilic layered surface assisted in the multi-layer deposition. PEDOT:PSS film preparation were similar to the earlier work[1]. Au electrodes were thermally evaporated as the counter electrode for both PEDOT:Tos and PEDOT:PSS structures. (Additive treatment of solvents in the nanopores cannot be accomplished due to residual or dewetting effects which may be improved further by infiltration techniques.)

*(c) Methods*

*Electrical characterization:* Keithley 4200 SCS parameter analyzer was used for impedance measurements. Linkam LTS420E-PB4 Temperature Controlled Stage was used for temperature-dependent frequency response studies. The effects from external contributions like long cable/wire coupling and parasitic capacitances coming to play in high (~MHz) frequency conductance measurements were eliminated by proper impedance compensation originating from these factors and using low loss connectors. Dc transport characteristics was studied in temperature range of 10 K< T< 340 K using Quantum Design Physical Property Measurement System Model 6000. Electrical transport via single nanochannels was measured using Conducting Atomic Force Microscopy (CAFM). For current imaging, JPK Nanowizard 3 with Au coated cantilever tips (Thickness: 2 µm/ Width: 50 µm/ Length: 450 µm/, Resonance Frequency: 13 kHz/ Force constant: 0.2 N/m) was used in contact mode.

*Electrical noise measurement:* Electrical noise measurements of the nanopore filled PEDOT structures were performed using Agilent 35670A Digital Signal Analyzer (a low noise 16-bit fft spectrum analyzer) that sampled and stored voltage signal. A current to voltage preamplifier DLPCA 200 (Femto) was used to amplify the current signal from DUT. Grounding and isolation of the sample stage were done in steps similar to earlier reported results on noise spectroscopy of polymer transistors[4] in our laboratory. Linkam LTS420E-PB4 Temperature Controlled Stage was used for temperature-dependent noise measurements.

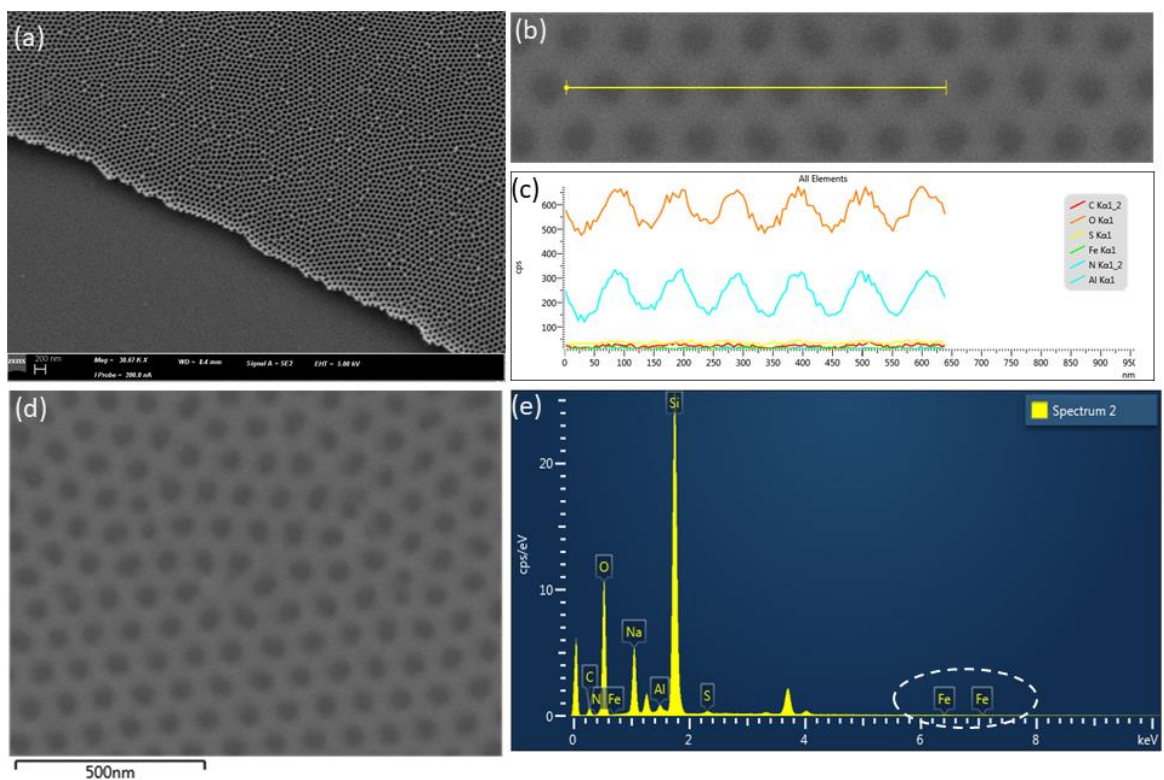

**Figure S1**: SEM image of (a) partially filled PEDOT:Tos in 50 nm AAO channels deposited on glass substrates and (b) a section of image (a) on which EDX along the yellow line is shown in (c). (d) Electron image and (e) its corresponding EDX image of the entire scan area shows absence of traces of Fe salt contents (marked in circle) on films. After every layer deposition, the films were washed thoroughly in ethanol to remove any trace amount of Fe content in the nanochannels.

## 2. Experiments showing complete filling of nanochannel with the polymer

The PEDOT:Tos coated AAO nanochannels on ITO are shown to completely fill the length of channels. After completion of the deposition methods including proper annealing of samples, clean scotch tapes are used to manually exfoliate layers of PEDOT:Tos/AAO from ITO surface. AFM imaging are performed on these films that adhere to the adhesive side of tape and also the residual PEDOT/AAO material on ITO. Few attempts of a quick tape exfoliation procedure show a partial removal of PEDOT/AAO from ITO surface leaving behind some PEDOT/AAO material on ITO. AFM scan on these surfaces proves the polymer filling midway along the length of nanochannels (**Figure S2(c)**). Whereas a complete exfoliation of PEDOT/AAO from ITO surface is possible with proper adhesion of tape for longer duration. The distinct topography and phase images of noncontact AFM scan (**Figure S2(d)**) on these adhesive tape-side show the polymer filled channels.

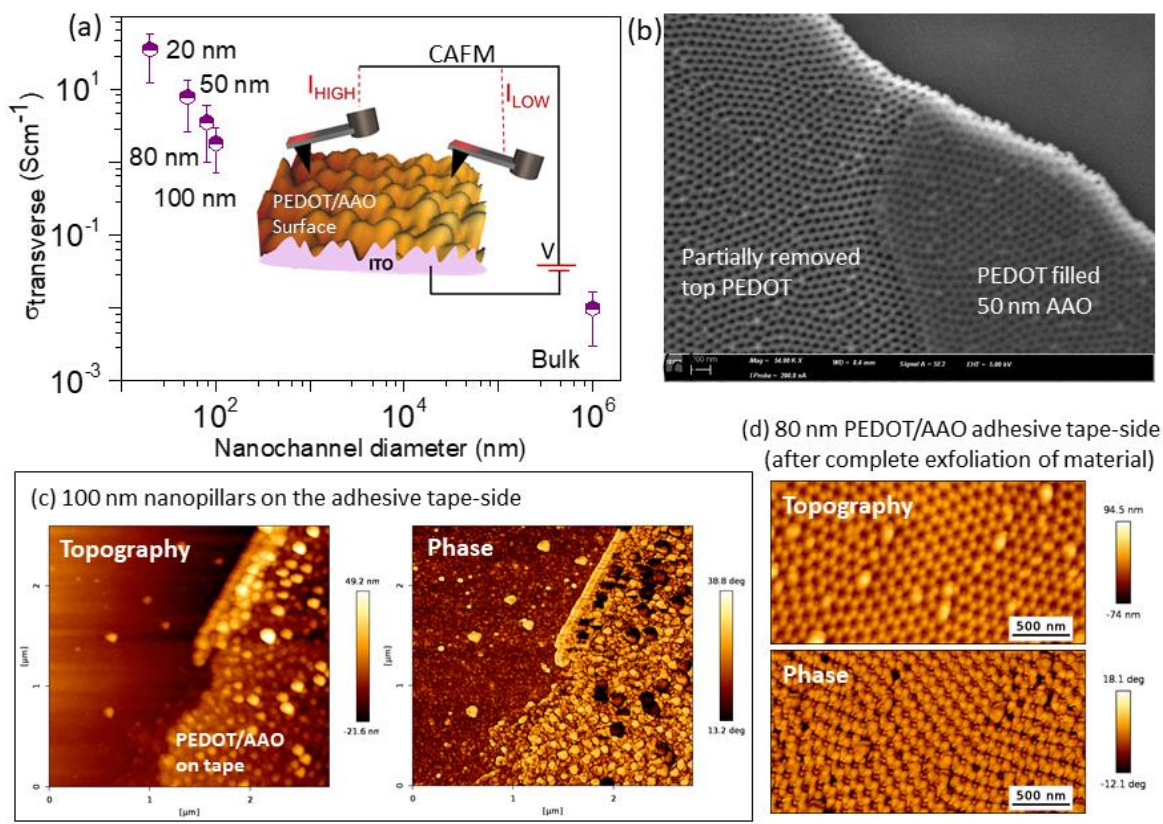

**Figure S2:** $\sigma_{transverse}$ versus PEDOT:Tos nanochannel diameters arrived from the results of CAFM current scans. (inset: schematic shows $I_{LOW}$ when the conducting tip encounters a nonporous alumina region and and $I_{HIGH}$ corresponds to that of PEDOT filled channels). (b)SEM image of PEDOT:Tos in 50 nm AAO on ITO, left section is exfoliated with tape to remove top polymer film and right section contains thin film of PEDOT/AAO. (c) Topography and phase images of partially peeled off PEDOT/AAO adhered on tape-side showing the exposed nanopillar-like PEDOT structures. (d) Topography and phase images of PEDOT/AAO on tape-side on complete removal of PEDOT/AAO from ITO surface.

*2.1 CAFM imaging on the PEDOT/AAO adhered to ITO (on partial exfoliation)*

Significant current magnitudes with contrasting nonporous and porous regions on the residual PEDOT/AAO on ITO after partial exfoliation of the material is direct evidence of conformal filling of polymers along the length of channels.

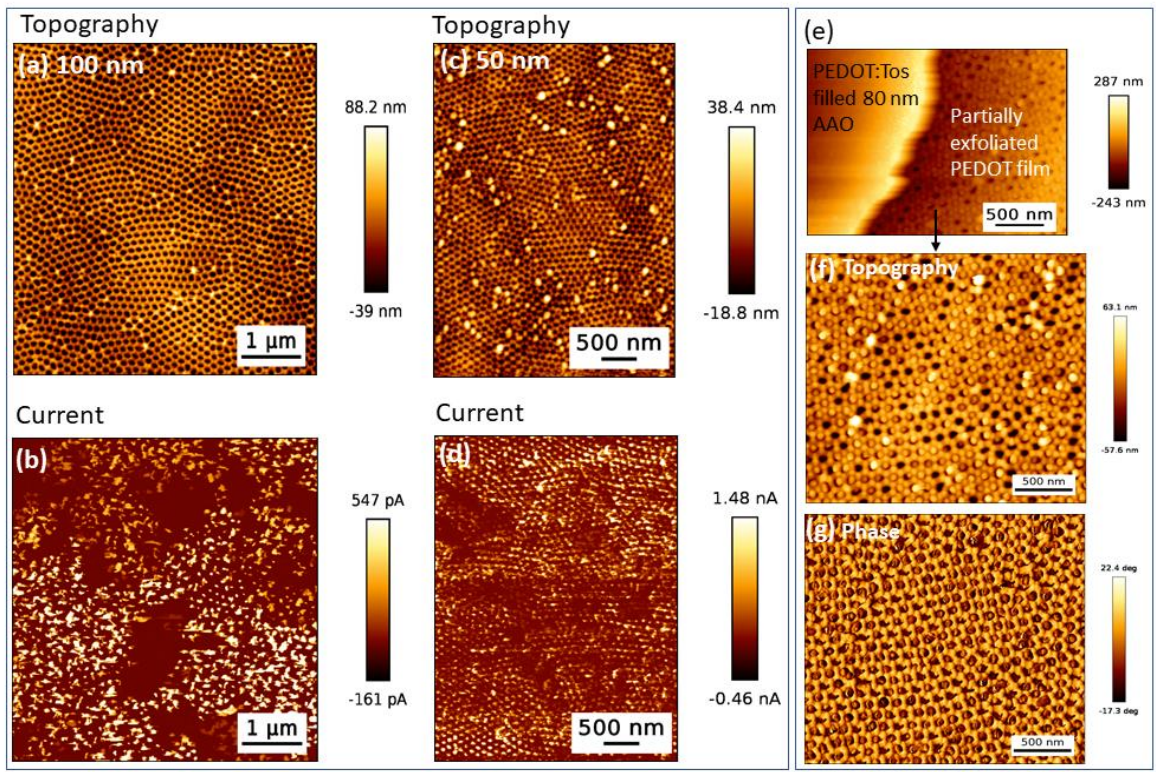

**Figure S2.1**: Contact mode current imaging: (a,c) Topography and (b,d) current profiles on residual 100 nm and 50 nm PEDOT/AAO surface adhered to ITO substrates, respectively after partial removal of the PEDOT/AAO material. (e) Topography of 80 nm PEDOT:Tos on ITO substrate, the marked right section is partially exfoliated that leaves exposed PEDOT on AAO template. The (f) topography and (g) phase images are shown for these exposed PEDOT on alumina.

The phase values in **Figure S2.1(g)** of PEDOT/AAO on ITO substrate clearly shows PEDOT in porous region and alumina in nonporous region. These noncontact mode topography and phase images on ITO substrates and the corresponding images on tape side (**Figure S2(d)**) distinctly indicates to the conformal filling of polymer in the nanochannels.

## 3. Noncontact mode surface morphology of PEDOT:Tos nanochannels

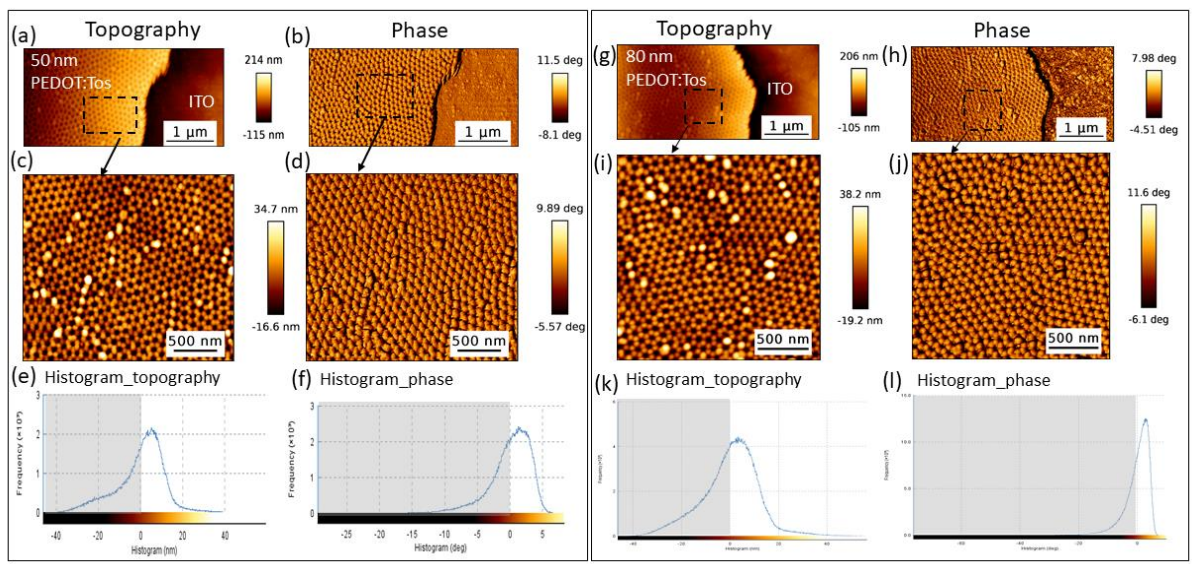

**Figure S3:** Noncontact surface morphology. (a,c) Topography and (b,d) phase across 50 nm PEDOT:Tos AAO/ITO interface over a scan area of 4.2 µm*4.0 µm and on PEDOT:Tos filled 50 nm AAO surface over a scan area of 1.5 µm *1.5 µm, respectively.(e,f) Histogram plots of topography and phase, respectively of scan area in (c,d). (g,i) Topography, (h,j) phase and (k,l) histogram plots are the corresponding images of that across 80 nm PEDOT:Tos.

The moderately contrasting phase values (b,d,h,j) are indicative of the same material present throughout the alumina template of 50 nm and 80 nm channels. The shift in the peak positions of the histogram in (e,k) topography and the (f,l) phase images clearly show a proportionate amount of negative region in topography as compared to the small amount of negative region in phase image. This can be observed as AFM tip encountering the polymer in most of the region of nanochannel.

## 4. Contact mode current imaging of PEDOT:Tos nanochannels

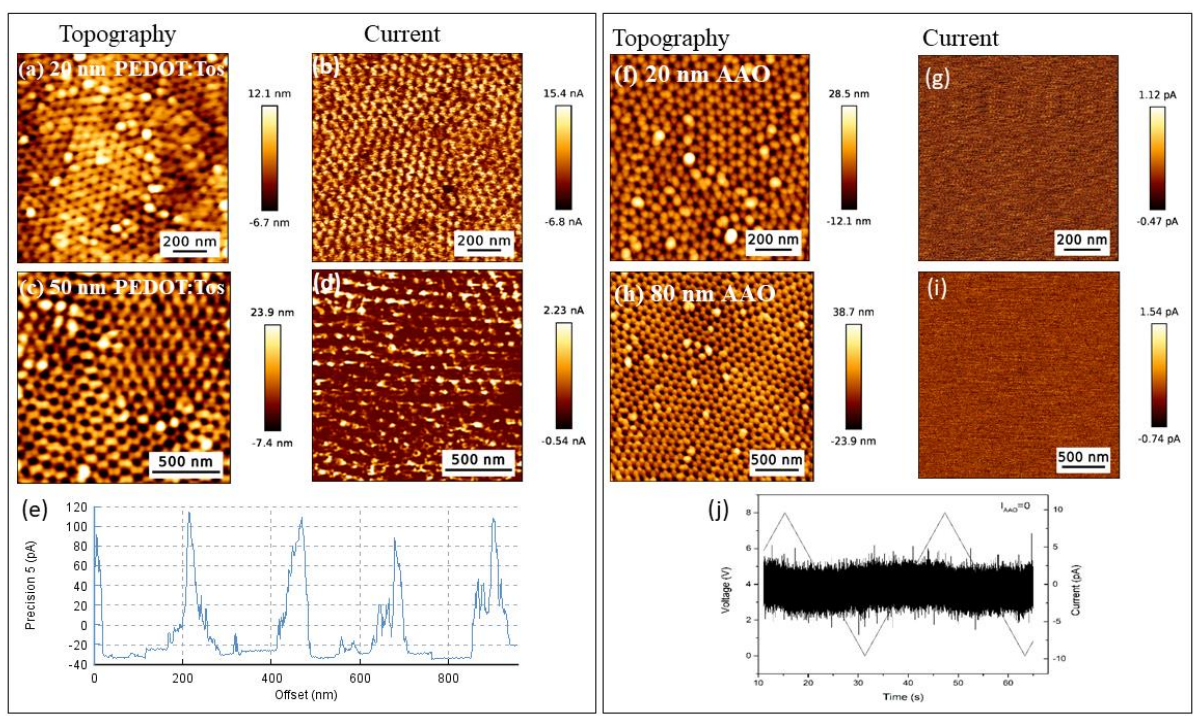

**Figure S4**: Single nanochannel electrical current mapping. (a,c) Topography and (b,d) current profiles of 20 nm and 50 nm PEDOT:Tos nanochannels on ITO, respectively. The PEDOT filled porous regions corresponds to the high current values providing a contrast with the low background current on nonporous AAO region. (e) Line scan of current profile. (f,h) and (g,i) are the topography and current profile, respectively of 20 nm and 50 nm bare AAO (no polymer

filled) on ITO, (j) V-t graph of a single nanochannel upon application of triangular bias voltage showing no current.

Figure S4 (g,i,j) confirms there is no current contribution from alumina region and the high current magnitudes in (b,d,e) are solely from the PEDOT:Tos in the nanopores.

## 5. $\sigma_{dc}(T)$ for PEDOT:Tos nanochannels (linear scale)

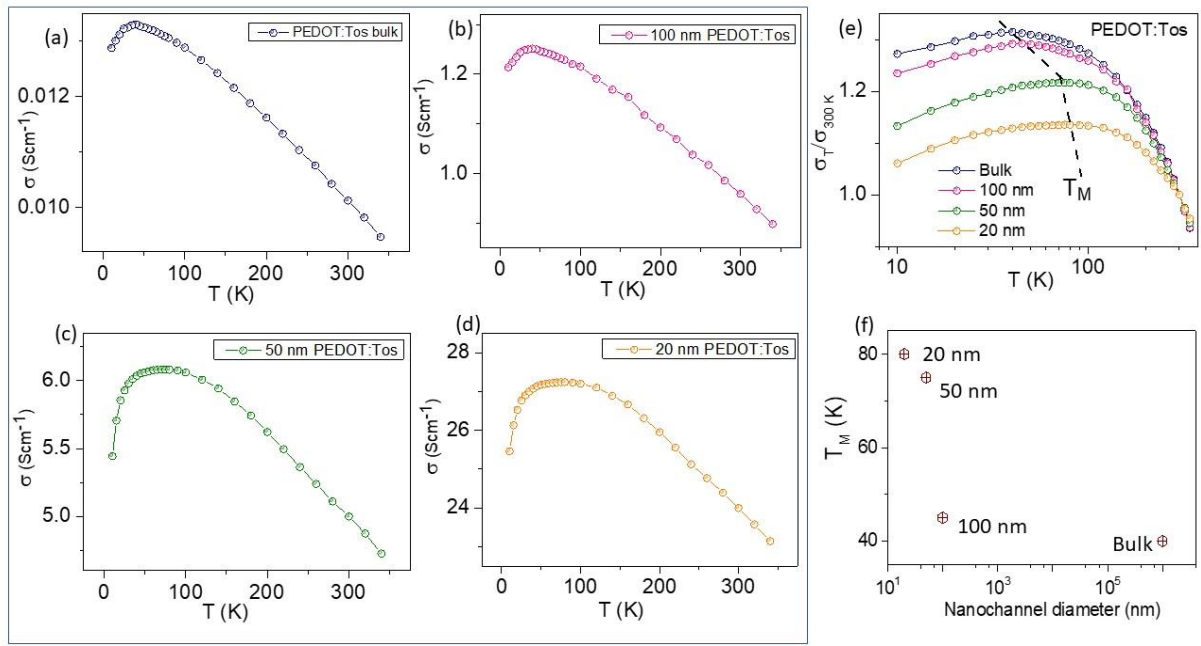

**Figure S5**: Temperature dependence of $\sigma_{dc}$ of (a) bulk transverse thin film, (b) 100 nm, (c) 50 nm, (d) 20 nm PEDOT:Tos nanochannels in the range of 10 K<T<340 K (all the graphs are shown in linear y-axis. (e) $\sigma_T / \sigma_{300K}$ versus T for all the confined PEDOT:Tos samples and the transition temperature $T_M$ is marked as shown. (f) $T_M$ plotted as a function of PEDOT:Tos nanochannel dimensions.

*5.1 Reduced Activation energy*

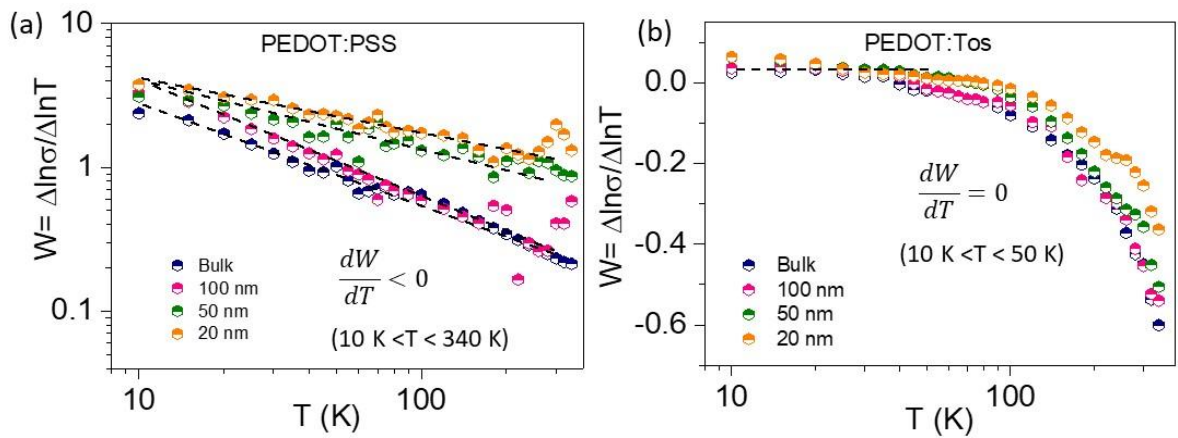

**Figure S5.1**: Reduced activation energy, W of (a) PEDOT:PSS and (b) PEDOT:Tos nanochannels.

Any existing disorder induced critical behavior near metal-insulator transition can be estimated from the reduced activation energy, W = $\Delta\ln\sigma/\Delta\ln T$ which is plotted against T in **Figure S5.1**. Confined PEDOT:PSS in all channel dimensions show negative slope of W(T) in temperature range 10 K< T< 340 K, indicating nonmetallic signatures.[5-7] All the confined and bulk PEDOT:Tos samples show nearly constant temperature coefficient of W(T) for T<$T_M$ and reflects a critical region in 10 K <T <$T_M$. It is to be noted that d$\sigma$/dT<0 for T>$T_M$. These characteristic transport behavior indicate that the PEDOT:Tos systems are in metallic regime.

## 6. PEDOT crystalline domains

The PEDOT crystalline domains are surrounded by amorphous regions in PEDOT:Tos films.

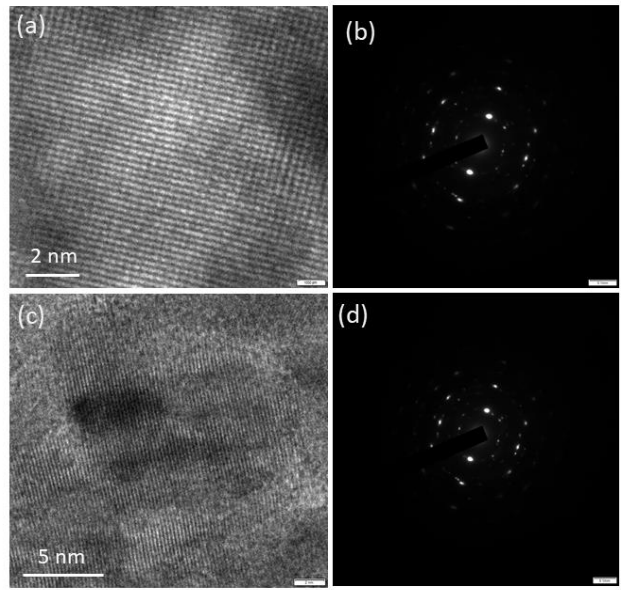

**Figure S6**: (a,c) HRTEM images of PEDOT:Tos films and (b,d) shows corresponding diffraction patterns.

## 7. Ac conductivity σ(ω,T) of PEDOT:Tos nanochannels

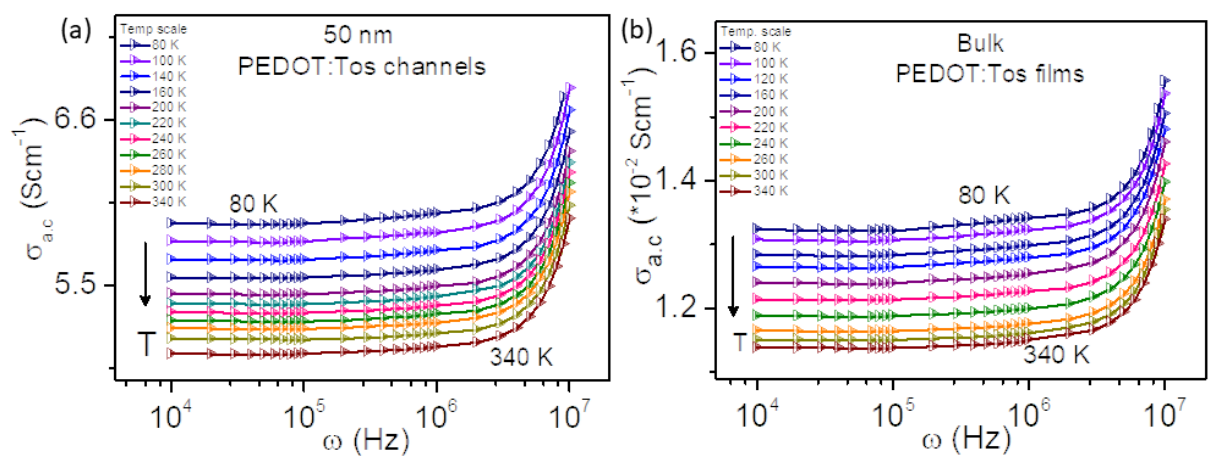

**Figure S7.** Ac conduction: $\sigma_{ac}$ versus frequency of (a) 50 nm and (b) bulk (thin fims) of PEDOT:Tos in temperature range 80 K < T < 340 K.

## 8. Electrical noise analysis

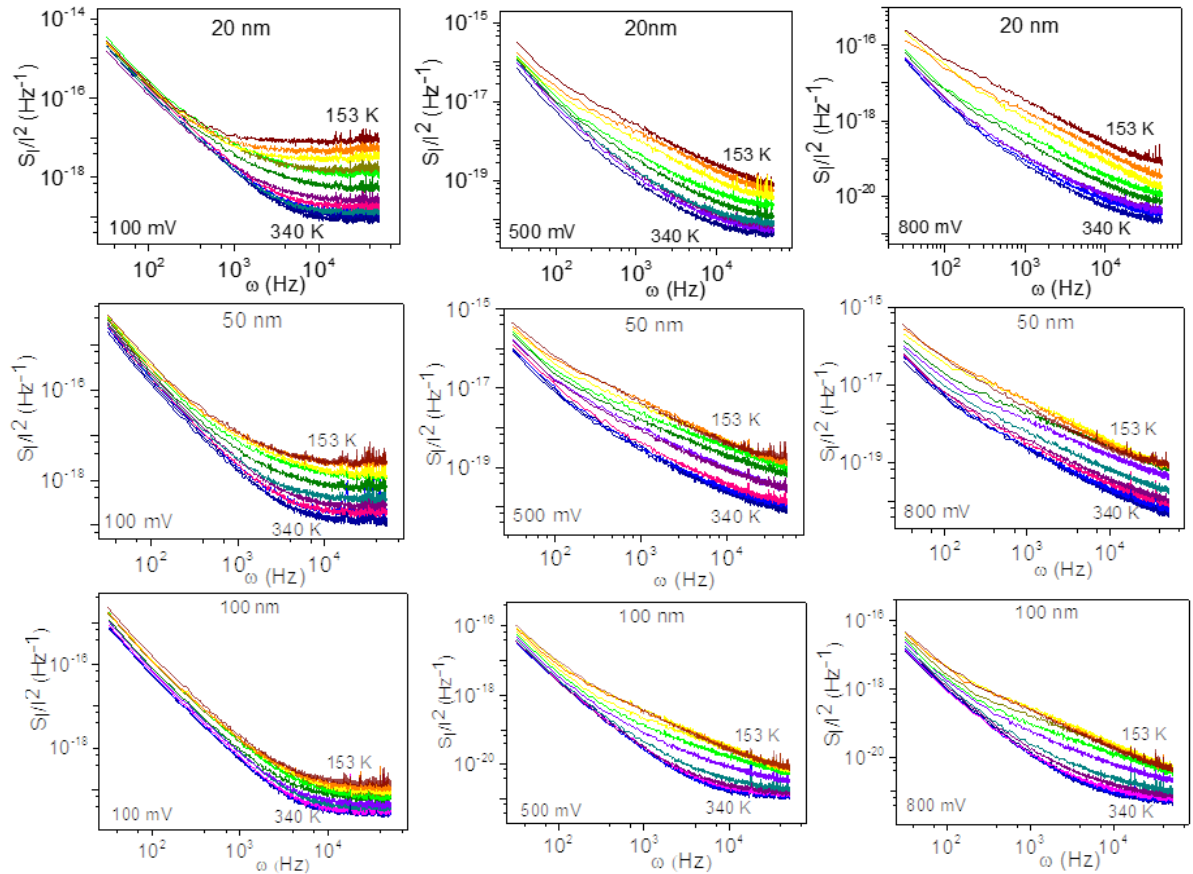

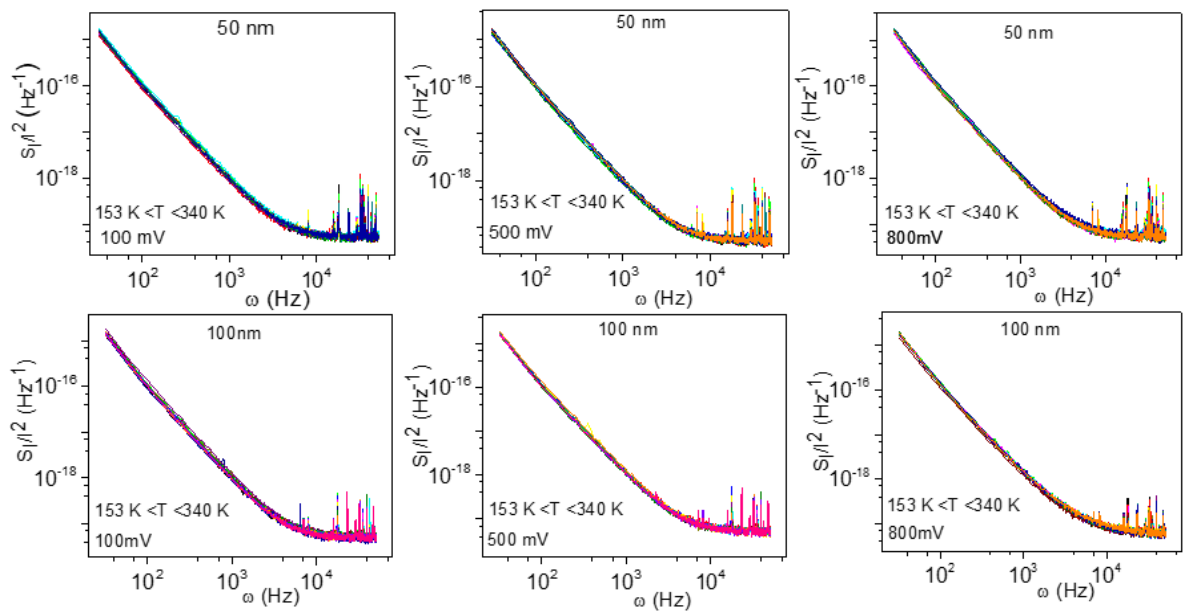

**Figure S8:** Noise studies in 50 nm and 100 nm AAOs for (a) PEDOT:PSS and (b) PEDOT:Tos nanochannels over temperature range 153 K $<$T $<$ 340 K and different voltage bias conditions of 100 mV, 500 mV and 800 mV.

Wide variation of noise spectral density over temperature range 153 K $<$ T $<$ 340 K in PEDOT:PSS nanochannels are observed in **Figure S8a** as compared to bias and T independent behavior in PEDOT:Tos nanochannels (**Figure S8b**).